\documentclass[letterpaper]{JHEP3}
\usepackage{amsfonts}

\usepackage{graphicx}
\usepackage{epsfig}

\newcommand{\beq}{\begin{equation}}
\newcommand{\eeq}{\end{equation}}
\newcommand{\beqa}{\begin{eqnarray}}
\newcommand{\eeqa}{\end{eqnarray}}
\newcommand{\beqar}{\begin{eqnarray*}}
\newcommand{\eeqar}{\end{eqnarray*}}

\newcommand{\norm}[1]{\raise.3ex\hbox{:}#1\raise.3ex\hbox{:}}

\abstract{We discuss the Reissner--Nordstr\"om--de Sitter black holes in the context
of dS/CFT correspondence by using static and planar coordinates. The boundary
stress tensor and the mass of the solutions are computed. Also, we investigate
how the RG flow is changed for different foliations. The Kastor--Traschen 
multi-black hole solution is considered as well as AdS counterparts of these configurations. In particular, we find that in planar coordinates the black holes appear like punctures in the dual boundary theory.}
 \preprint{hep-th/0310273\\
\newline \small \hfill FREIBURG-THEP-03-14}

\title{Reissner--Nordstr\"om--de Sitter black hole, planar coordinates and
dS/CFT}
\author{Dumitru Astefanesei,$^1$\thanks{%
E-mail: \texttt{astefand@hep.physics.mcgill.ca}} \ Robert B. Mann$^{2}$%
\thanks{%
E-mail: \texttt{mann@avatar.uwaterloo.ca}} \ and Eugen Radu$^3$\thanks{%
E-mail: \texttt{radu@newton.physik.uni-freiburg.de}} \\
$^{1}$Department of Physics, McGill University Montr\' eal, Qu\' ebec H3A
2T8, Canada\\
$^{1,2}$Perimeter Institute for Theoretical Physics, Ontario N2J 2W9, Canada%
\\
$^{2}$Department of Physics, University of Waterloo Waterloo, Ontario N2L
3G1, Canada\\
$^{3}$Physikalisches Institut, Albert-Ludwigs-Universit\"at Freiburg,
D-79104, Freiburg, Germany}
\begin{document}

\section{Introduction}

There are a number of observational as well as theoretical reasons that
motivate, at least partially, the recent interest in asymptotically de
Sitter (dS) spacetimes. The observational evidence accumulated in recent
years (see, $e.g.$, ref.~\cite{data}), seems to favour the idea that the
physical universe has an accelerated expansion. The most common explanation
is that the expansion is driven by a small positive vacuum energy
(cosmological constant, $\Lambda(G\hbar/c^3)\approx 10^{-123}$), implying
the spacetime is asymptotically dS. Furthermore dS spacetime plays a central
role in the theory of inflation (the very rapid accelerated expansion in the
early universe), which is supposed to solve the cosmological flatness and
horizon puzzles.

dS spacetime is the maximally symmetric Lorentzian space ($i.e.$ the number
of Killing vector fields is the same as for flat spacetime) with constant
positive curvature. In $D$-dimensions the symmetry group is $SO(1,D)$ and
the topology is $R\times S^{D-1}$. The existence of compact Cauchy surfaces
(spatial slices), easier seen from the topology of dS spacetime, makes
unclear the procedure to define conserved charges. The causal structure of
dS spacetime is such that inertial observers are surrounded by cosmological
horizons. The expansion is so rapid that no signal originating outside any
observer's horizon will ever reach her; equivalently there is a null surface
that cannot be crossed by any material particle or signal. If an observer
does not have access to a part of the spacetime, then she will attribute an
entropy to the gravitational field because of the microscopic degrees of
freedom that are hidden.

A cosmological horizon is a characteristic of many spacetimes having a
positive cosmological constant. The next logical step is to define an energy
for such a spacetime in order to provide consistent thermodynamic
relationships, a problematic task in general relativity. One of the most
fruitful approaches in computing conserved quantities has been to employ the
quasilocal formalism \cite{brown}. The basic idea here is to enclose a given
region of spacetime with some surface, and to compute all relevant
(conserved and/or thermodynamic) quantities with respect to that surface %
\cite{bcm,booth}. For a spacetime that is either asymptotically flat or
asymptotically anti de Sitter it is possible to extend the quasilocal
surface to spatial infinity without difficulty, provided one incorporates
appropriate boundary terms in the action to remove divergences \cite{bcm}-%
\cite{kls}. For pure and asymptotically de Sitter spacetimes inside the
cosmological horizon (where the Killing vector is timelike) computations of
conserved charges and actions/entropies have been carried out \cite%
{Dehghani1}.

The microphysical statistical origin of cosmological horizon entropy may
well be associated with a holographic dual theory. Inspired by the
celebrated anti-de Sitter (AdS)--conformal field theory (CFT) correspondence %
\cite{Maldacena:1997re}, some authors have conjectured a holographic duality
between quantum gravity in an asymptotically dS spacetime and a CFT living
on its boundary \cite{Hull:1998vg}-\cite{Strominger:2001pn}. However, in
spite of considerable recent activity, the precise nature of the proposed
dS/CFT correspondence remains vague, with numerous conceptual issues
remaining to be addressed. First, there is no satisfactory realization of dS
space in string theory. Also, unlike in AdS, in asymptotically dS spacetimes
there is no asymptotic Killing vector that is globally timelike.
Consequently the Hamiltonian is not positive definite and there is no
spatial infinity --- the asymptotic regions are Euclidean surfaces (spheres)
at past and future temporal infinities \cite{Hawking}. The conjectured
dS/CFT correspondence introduces holographic screens at timelike past
infinity $\mathcal{I}^{-}$ and/or timelike future infinity $\mathcal{I}^{+}$
(see, $e.g$, ref.~\cite{banks} for a different approach --- the holographic
screens, at the cosmological horizons, are observer dependent). The theory
on the screen is necessarily a Euclidean CFT, with a scale that encodes the
dimension transverse to the screen.

Despite these complications, we have learned a lot about asymptotically dS
spacetimes in recent years. A novel method for renormalizing the
stress-energy of gravity that provides a measure of the gravitational mass
(and the boundary stress tensor) was proposed by Balasubramanian, de Boer
and Minic in ref.~\cite{Balasubramanian:2001nb} (referred to thereafter as
the BBM prescription). This method is analogous to the Brown--York
prescription in asymptotically AdS spacetimes: one supplements the
quasilocal formalism by including boundary counterterms that depend on the
curvature invariants. By generalizing the Gibbs--Duhem relation, a
definition of entropy outside the cosmological horizon was formulated in
ref.~\cite{Ghezelbash:2002ab}. This formalism was recently used to study the
thermodynamic properties of four-dimensional Kerr--Newman--dS black holes in
stationary coordinates \cite{Dehghani2}.

Because of its high degree of symmetry, dS space has a simple form in a
large number of coordinate systems. There is a static frame centered on each
observer (timelike geodesic) in dS. When a black hole exists, there is still
a static frame centered about the black hole. Since different
parametrizations emphasize different features, it is of interest to study
physics in dS space in alternative coordinate systems. For example, a
multi-black hole solution in dS can be formulated in cosmological
coordinates --- one cannot build a static frame for this situation.
Moreover, by choosing different foliations of the spacetime one can describe
boundaries that have different topologies and geometries (metrics),
affording study of the CFT on different backgrounds. Specifically, we find
additional Casimir-type contributions to the total energy depending on the
slicing topology in accord with the expectations from quantum field theory
in curved space \cite{Shiromizu2002}.

In what follows we will use two parametrizations of dS$_{n+2}$ (with $n>1$).
The first we will refer to as static coordinates 
\[
ds^{2}=(1-H^{2}R^{2})^{-1}dR^{2}+R^{2}d\Omega _{n}^{2}-(1-H^{2}R^{2})dT^{2}, 
\]
with $d\Omega _{n}^{2}=\omega _{ab}dx^{a}dx^{b}=d\theta ^{2}+\sin ^{2}\theta
d\Omega _{n-1}^{2}$ the unit metric on $S^{n}$. Throughout the paper we set $%
c=G=1$; also, the indices $\{a,b,\ldots \}$ will indicate the angular
coordinates and $\{i,j,\ldots \}$ will indicate the intrinsic coordinates of
the boundary metric. The cosmological constant is $\Lambda =n(n+1)H^{2}/2$.
This parametrization is particularly useful, since the metric looks very
simple and is time independent. The existence of a cosmological horizon at $%
R=1/H$, with all thermodynamical properties, is here manifest \cite%
{Gibbons:mu}.

The second set, which we will refer to as planar coordinates ($i.e.$ spatial
slices are flat) --- or cosmological or inflationary coordinates --- is
given by 
\[
ds^{2}=e^{2Ht}(dr^{2}+r^{2}d\Omega _{n}^{2})-dt^{2}. 
\]
In these isotropic coordinates, the expansion of the universe is explicit,
and we can address questions relevant for a cosmology with a period of
inflation. 
The relationship between these coordinate patches and their Penrose diagrams
are presented in ref. \cite{Hawking}.

Similar to the AdS case, black holes in dS space are expected to provide
crucial information in understanding the dS/CFT correspondence. Most of the
studies on cosmological black hole spacetimes make use of the static
coordinates. It is less well-known that (electrovacuum-) black hole physics
can also be discussed using a planar coordinate system. Although such
metrics have explicit time-dependence, they also have a Killing vector that
is equivalent to the $\partial /\partial T$ in static coordinates.
Alternatively, the ``static'' metric is time-dependent outside of the
cosmological horizon. Either way --- in planar or static coordinates --- it
is a time dependent spacetime since there is no global timelike isometry.

A discussion of the relationship between planar and static coordinates for a
Schwarzschild--de Sitter black hole, from the viewpoint of the dS/CFT
correspondence conjecture, can be found in refs.~\cite{klemm,
Danielsson:2001wt, Shiromizu:2001bg}. The simplest generalization of the
Schwarzschild black hole is to include a $U(1)$ field in the theory. For a
negative $\Lambda$, the corresponding Einstein--Maxwell (EM) solutions have
proven useful in understanding various aspects of AdS/CFT \cite%
{Chamblin:1999tk}. The relevance of Reissner--Nordstr\"om--de Sitter (RNdS)
black holes within the dS/CFT correspondence has been discussed by a number
of authors, using a static coordinate system. 
However, the RNdS solution also admits a simple form in a planar coordinate
system \cite{Brill:1993tw}, which allows for a multi-black hole
generalization given by Kastor and Traschen in ref.~\cite{Kastor:1992nn}
(referred to as the KT solution). This solution describes an arbitrary
number of charged black holes which, due to the presence of a cosmological
constant, are in motion with respect to one another. In the following we
re-examine these solutions in the context of the dS/CFT correspondence.
Using the planar coordinate system, we find that the black holes appear as
punctures in the dual theory, corresponding to divergences in the stress
tensor of the CFT.

The remainder of our paper is organized as follows: we start by reviewing in
Section 2 the counterterm formalism for asymptotically dS spacetimes. In
Subsection 2.1 we discuss the RNdS solution in static coordinates, compute
its entropy working in a grand canonical ensemble and comment on its
relationship to the Cardy--Verlinde formula of the \textit{hypothetical}
dual CFT. In Subsection 2.2 we investigate the RNdS black hole in planar
coordinates and in Section 3 we describe the multi-black hole generalization
in the context of the dS/CFT correspondence. A discussion of RG flow and the
c-function for the RNdS solution is given in Section 4. Using a double
analytical continuation, we build in Section 5 new time-dependent
backgrounds with a negative cosmological constant. We conclude in Section 6
with a discussion of our results.


\section{Counterterm method and conserved charges}

The mass and boundary stress-tensor as well as the thermodynamic quantities
of a RNdS solutions are computed here by applying for the EM system the
general formalism presented in refs.~\cite{Ghezelbash:2002ab,Clarkson2003c}
(see, also, ref.~\cite{Balasubramanian:2001nb}).

We start by considering the path integral 
\begin{equation}
\left\langle g_{2},\Phi _{2},S_{2}|g_{1},\Phi _{1},S_{1}\right\rangle =\int
D \left[ g,\Phi \right] \exp \left( iI\left[ g,\Phi \right] \right),
\label{PI1}
\end{equation}
which represents the amplitude to go from a state with metric and matter
fields $\left[ g_{1},\Phi _{1}\right] $ on a surface $S_{1}$ to a state with
metric and matter fields $\left[ g_{2},\Phi _{2}\right] $ on a surface $%
S_{2} $. \ The quantity $D\left[ g,\Phi \right] $ is a measure on the space
of all field configurations and $I\left[ g,\Phi \right] $ is the action
taken over all fields having the given values on the surfaces $S_{1}$ and $%
S_{2}$.

For asymptotically dS spacetimes we replace the surfaces $S_{1},S_{2}$ with
histories $H_{1},H_{2}$\ that have spacelike unit normals and are surfaces
that form the timelike boundaries of a given spatial region. The amplitude (%
\ref{PI1}) then is 
\begin{equation}
\left\langle g_{2},\Phi _{2},H_{2}|g_{1},\Phi _{1},H_{1}\right\rangle =\int
D \left[ g,\Phi \right] \exp \left( iI\left[ g,\Phi \right] \right)
\label{PI2}
\end{equation}%
and describes quantum correlations between differing histories $\left[
g_{1},\Phi _{1}\right] $ and $\left[ g_{2},\Phi _{2}\right] $\ of metrics
and matter fields, with the modulus squared of the amplitude yielding the
correlation between two histories. Spacelike tubes at some initial and final
times join the surfaces $H_{1}, H_{2}$, so that the boundary and interior
region are compact. As these times approach past and future infinity one
obtains the correlation between the complete histories, given by summing
over all metric and matter field configurations that interpolate between
them. The result does not depend on any special hypersurface between the
hypersurfaces $H_{1}$\ and $H_{2}$.

We decompose the action into the distinct parts 
\begin{equation}
I=I_{B}+I_{\partial B}+I_{ct}  \label{actiongeneral}
\end{equation}%
where the bulk ($I_{B}$) and boundary ($I_{\partial B}$) terms are the usual
ones, given by 
\begin{eqnarray}
I_{B} &=&\frac{1}{16\pi }\int_{\mathcal{M}}d^{n+2}x~\sqrt{-g}\left(
R-2\Lambda +\mathcal{L}_{M}(\Phi )\right)  \label{actionbulk} \\
I_{\partial B} &=&-\frac{1}{8\pi }\int_{\partial \mathcal{M}^{\pm }}d^{n+1}x~%
\sqrt{h^{\pm }}K^{\pm }  \label{actionboundary}
\end{eqnarray}%
where $\partial \mathcal{M}^{\pm }$ represents future/past infinity, and $%
\int_{\partial \mathcal{M}^{\pm }}=\int_{\partial \mathcal{M}^{-}}^{\partial 
\mathcal{M}^{+}}$ represents an integral over a future boundary minus a past
boundary, with the respective metrics $h^{\pm }$ and extrinsic curvatures $%
K^{\pm }$. The quantity in (\ref{actionbulk}) $\mathcal{L}_{M}(\Phi )=-F^{2}$
is the Lagrangian for the Maxwell field. The bulk action is over the $\left(
n+2\right) $--dimensional manifold $\mathcal{M}$, and the boundary action is
the surface term necessary to ensure well-defined Euler--Lagrange equations.

For an asymptotically dS spacetime, the boundary $\partial \mathcal{M}$ will
be a union of Euclidean spatial boundaries at early and late times. The
dS/CFT-inspired counterterm action $I_{ct}$ is defined on these boundaries
as suggested by analogy with the AdS/CFT correspondence, and its terms
depend only on geometric invariants of these spacelike surfaces. It is
universal in both the asymptotic AdS and dS cases \cite%
{kls,Ghezelbash:2002ab}, and can be generated by an algorithmic procedure
without reference to a background metric. In the dS case one obtains \cite%
{Ghezelbash:2002ab} 
\begin{eqnarray}
I_{ct} &=&-\frac{1}{8\pi }\int d^{n+1}x\sqrt{\gamma }\left\{ -\frac{n}{\ell }%
+\frac{\ell \mathsf{\Theta }\left( n-2\right) }{2(n-1)}\mathsf{R}-\frac{\ell
^{3}\mathsf{\Theta }\left( n-4\right) }{2(n-1)^{2}(n-3)}\left( \mathsf{R}%
_{ij}\mathsf{R}^{ij}-\frac{n+1}{4n}\mathsf{R}^{2}\right) \right.  \nonumber
\\
&&-\frac{\ell ^{5}\mathsf{\Theta }\left( n-6\right) }{(n-1)^{3}(n-3)(n-5)}%
\left( \frac{3n+5}{4n}\mathsf{RR}^{ij}\mathsf{R}_{ij}-\frac{(n+1)(n+3)}{%
16n^{2}}\mathsf{R}^{3}\right.  \nonumber \\
&&\left. -2\mathsf{R}^{ij}\mathsf{R}^{kl}\mathsf{R}_{ijkl}\left. -\frac{n+1}{%
4n}\nabla _{k}\mathsf{R}\nabla ^{k}\mathsf{R}+\nabla ^{k}\mathsf{R}%
^{ij}\nabla _{k}\mathsf{R}_{ij}\right) +\ldots \right\} ,  \label{actionct}
\end{eqnarray}%
with $\mathsf{R}$ the curvature of the induced metric $h_{ij}$ and, to
assure consistency with previous works, we noted $H=1/l$. The step-function $%
\mathsf{\Theta }\left( x\right) $ is unity provided $x>0$ and vanishes
otherwise, thereby ensuring that the series only contains the terms
necessary to cancel divergences and no more. In four ($n=2$) dimensions, for
example, only the first two terms appear, and only these are needed to
cancel divergent behavior in $I_{B}+I_{\partial B}$ near past and future
infinity.

The boundary metric can be written, at least locally, in ADM-like form 
\begin{equation}
d\hat{s}^{\pm 2}=h_{ij}^{\pm }d\hat{x}^{\pm i}d\hat{x}^{\pm j}=N_{\tau
}^{\pm 2}d\tau ^{2}+\sigma _{ab}^{\pm }\left( d\phi ^{\pm a}+N^{\pm a}d\tau
\right) \left( d\phi ^{\pm b}+N^{\pm b}d\tau \right) ,  \label{hmetric}
\end{equation}%
where $N_{\tau }$ and $N^{a}$ are the lapse function and the shift vector
respectively and the $\phi ^{a}$ are the intrinsic coordinates on the closed
surfaces $\Sigma $. It is clear that $\nabla _{\mu }\tau $ is a spacelike
vector field that is the analytic continuation of a timelike vector field,
the boundary metric(s) being euclidean (spacelike tube(s)) for
asymptotically dS spacetimes.

Varying the action with respect to the boundary metric $h_{ij}$, we find the
boundary stress-energy tensor for gravity 
\begin{equation}
T^{\pm ij}=\frac{2}{\sqrt{h^{ \pm}}} \frac{\delta I}{\delta h_{ij}^{\pm}}
\end{equation}%
(its explicit expression is given in ref.~\cite{Ghezelbash:2002ab}). In this
approach, the conserved quantities associated with a Killing vector $\xi
^{\pm i }$ are given by 
\begin{equation}
\mathfrak{Q_{\xi}}{}^{\pm }=\oint_{\Sigma ^{\pm }}d^{n}\phi ^{\pm }\sqrt{%
\sigma ^{\pm }}n^{\pm i}T_{ij}^{\pm }\xi ^{\pm j} ,  \label{Qcons}
\end{equation}%
where $n^{\pm i}$ is an outward-pointing unit vector, normal to surfaces of
constant $\tau$. Here, $\xi ^{i}$ needs not be a bulk Killing vector; the
quantity $\mathfrak{Q}_{\xi }$ will be conserved if $\xi ^{i}$ is a Killing
vector only on the boundary. Physically, this means that a collection of
observers, on the hypersurface with the induced metric $h_{ij}$, would all
measure the same value of $\mathfrak{Q}_{\xi }$ provided this surface has an
isometry generated by $\xi ^{i}$. For any two histories the value of $%
\mathfrak{Q}$ is the same for each. This surface does not enclose anything %
\cite{booth}; rather it is the boundary of the class of histories that
interpolate between $H_{1}$\ and $H_{2}$. Consequently $\mathfrak{Q}$ is not
associated with the class of histories that it bounds, but rather only with
this boundary \cite{Clarkson2003c}.

The conserved mass is defined to be 
\begin{equation}
\mathfrak{M}{}^{\pm }=\oint_{\Sigma ^{\pm }}d^{n}\phi ^{\pm }\sqrt{\sigma
^{\pm }}N_{\tau}^{\pm }n^{\pm i}n^{\pm j}T_{ij}^{\pm } ,  \label{Mcons}
\end{equation}%
provided $\partial /\partial \tau$ is a Killing vector on $\Sigma$; it is a
function of the cosmological time $\mathcal{T}$. As $\mathcal{T}$ approaches
positive or negative infinity, there is at least the notion of a conserved
total mass ${}\mathfrak{M}^{\pm }$ for dS spacetime, since all
asymptotically de Sitter spacetimes have an asymptotic isometry generated by 
$\partial /\partial \tau$.

An evaluation of the path integral may be carried out along the lines
described in ref. \cite{Clarkson2003c}. Since the action is in general
negative definite near past and future infinity (outside of a cosmological
horizon), we analytically continue the coordinate orthogonal to the
histories to complex values by an anticlockwise $\frac{\pi }{2}$-rotation of
the axis normal to them. This generally imposes, from the regularity
conditions, a periodicity $\beta$ of this coordinate, which is the analogue
of the Hawking temperature outside the cosmological horizon.

This renders the action pure imaginary, yielding a convergent path integral 
\begin{equation}
Z^{\prime }=\int e^{+\hat{I}}  \label{Partitionaction}
\end{equation}%
since $\hat{I}<0$. In the semi-classical approximation this will lead to $%
\ln Z^{\prime }=+I_{cl}$. For a grand canonical ensemble with fixed
temperature and fixed chemical potentials $C_i$ and conserved charges $\mu_i$
(which are $\Phi$ and $\mathbf{Q}$ for a RNdS solution) we can write 
\begin{equation}  \label{W}
W= \mathfrak{U}-TS -\mu_i C_i,
\end{equation}%
where $W$ is the grand canonical (Gibbs) potential and $T=1/\beta$. For a
converging partition function, we have $W=I_{cl}/\beta$ and thus we find for
the entropy of the system 
\begin{equation}  \label{S}
S =\beta (\mathfrak{U}-\mu_i C_i)-I_{cl}.
\end{equation}

\subsection{Static coordinates}

We now consider a charged black hole asymptotically dS spacetime in a static
coordinate system. The corresponding line element reads 
\begin{eqnarray}
ds^{2} &=&\frac{dR^{2}}{F(R)}+R^{2}d\Omega _{n}^{2}-F(R)dT^{2},  \label{rnds}
\\
F(R) &=&1-\frac{2M}{R^{n-1}}+\frac{Q^{2}}{R^{2(n-1)}}-H^{2}R^{2}.  \nonumber
\end{eqnarray}
and gives a solution of the EM equations for a (pure electric) gauge
potential 
\begin{eqnarray}  \label{A}
A=A_TdT=\Big(\sqrt{\frac{n}{2(n-1)}}\frac{ Q}{R^{n-1}}+\Phi \Big) dT,
\end{eqnarray}
where $\Phi$ is a constant (to be fixed below). In the above expressions, $M$
and $Q$ are constants proportional to the gravitational mass $\mathfrak{M}$
and the total electric charge $\mathbf{Q}$, respectively.

A discussion of this solution appeared in refs.~\cite%
{Brill:1993tw,Romans:1991nq}. Here we briefly review its basic properties.
The metric has a curvature singularity at the origin $R=0$. In general,
there are three kinds of Killing horizon at the radii where $F(R)$ vanishes.
Of interest are the outer black hole horizon at $R=R_{+}$ and the
cosmological horizon $R=R_{c}\leq 1/H$ corresponding to the largest root of $%
F(R)$. The two horizons are not in thermal equilibrium because the time
periods in the Euclidean section required to avoid a conical singularity at
both do not match in general. However, there are two families of RNdS
solutions for which the black hole and cosmological horizon temperatures do
match: the ``lukewarm'' black hole and charged Nariai black hole (see refs.~%
\cite{{ross},{bousso}} for a detailed discussion).

The lukewarm solutions are characterized by the condition $Q^{2}=M^{2}$,
being stable end-points of the evaporation process. In flat space, a black
hole evaporates until it becomes extremal. An extremal black hole is in
thermal equilibrium with radiation at any temperature. In dS spacetime the
situation is different. That is, a black hole (with $Q^{2}<M^{2}$) cannot
become arbitrarily cold --- the black hole will be stabilized by the
Gibbons-Hawking radiation from the cosmological horizon. If $Q^{2}>M^{2}$
the black holes are colder than the lukewarm solution and will absorb
radiation from the cosmological horizon. As $Q^{2}$ increases relative to $%
M^{2}$ the inner and the outer black hole horizons eventually coincide, and
the extremal (or ``cold'') RNdS black hole is obtained.

As the parameter $M$ increases (relative to $|Q|$, with $M>0$), the outer
black hole and cosmological horizons move closer together. The charged
Nariai solution is obtained when these horizons coincide at $%
R_{h}=R_{c}=R_{+}$; this is the largest charged asymptotically dS black
hole. For a given $Q$, the charged Nariai black hole has the maximal mass $%
M_N$. Solving $F(R)=F^{\prime}(R)=0$ (so that $R_{h}$ is a double root of $%
F(R)$) we obtain: 
\[
R_{h}^{n-1}=\frac{1}{2}M\left( n+1\right) \left( 1\pm \sqrt{1-\frac{4nQ^{2}}{%
\left( n+1\right) ^{2}M^{2}}}\right) 
\]
for the event horizon. The charged Nariai solution and lukewarm solution
meet for the critical value $Q^{2}=M_{c}^{2}$, where 
\[
M_{c}^{2}=\left( \frac{(n-1)^2}{H^{2}}\right) ^{n-1}\frac{1}{n^{2n}}\ . 
\]
A black hole in dS spacetime is obtained for a range of mass parameter $%
M_{min}\leq M\leq M_{N}$ (referred to as the ``undermassive'' case), where $%
M_{min}$ is the mass of extremal RNdS. If these limits are exceeded, the
metric (\ref{rnds}) describes a naked singularity.

The computation of the mass, action and entropy of a RNdS black hole is a
direct application of the method described in the previous section. We work
outside of the cosmological horizon, where $F(R)<0$. The topology of (\ref%
{rnds}), for large constant $R$, is an Euclidean cylinder $R\times S^{n}$
and $T$ is the coordinate along the cylinder. $\mathcal{I}^{\pm }$ are
located outside the future/past cosmological horizons, where $R$ is timelike
and $T$ is spacelike.

The gravitational mass/energy is the charge associated with the Killing
vector $\partial /\partial T$ --- now spacelike outside the cosmological
horizon. The total energy found by using the counterterm prescription is 
\begin{equation}
\mathfrak{U}=-\frac{\mbox{Area}(S^{n})}{8\pi } \ (nM-\frac{n}{H^{n-1}}\ 
\frac{\Gamma(\frac{2p-1}{2})}{2\sqrt{\pi}\Gamma(p+1)}\delta_{2p,n+1}),
\label{energy}
\end{equation}
for $p$ an integer, where $\mbox{Area}(S^{n})=2\pi ^{\frac{n+1}{2}}/\Gamma ({%
(n+1)}/2)$ is the area of a unit $n$--sphere. The second term is interpreted
as the Casimir energy in the context of dS/CFT correspondence. We have
explicitly checked this result up to $n=7$. This reduces to the expression
obtained in ref.~\cite{Ghezelbash:2002ab} when $Q=0$.

The total electric charge is computed by generalizing the methods of ref.~%
\cite{Jolien} to the de Sitter case. Working again outside of the
cosmological horizon, the electromagnetic field is 
\[
F_{RT}=-\sqrt{\frac{n(n-1)}{2}}\frac{Q}{R^{n}}, 
\]%
where $R$ is timelike and $T$ is spacelike. The induced metric $h_{\mu \nu
}=g_{\mu \nu }+u_{\mu }u_{\nu }$ projects the electromagnetic field on a
specific slice of the foliation. The electric field with respect to a slice $%
R=const.$ is $E_{i}=h_{i}^{\mu }F_{\mu \nu }u^{\nu }$ and the surface charge
density at future/past infinity is 
\[
\rho _{\mathbf{Q}}=\sqrt{\sigma ^{\pm }}n^{\pm i}E_{i}^{\pm }, 
\]%
where the vector $u^{\nu }=\left( \frac{1}{\sqrt{\left| F\left( R\right)
\right| }},0,\vec{0}\right) $ is normal to the induced metric $h_{\mu \nu }$
and $n^{i}=\left( 0,\sqrt{\left| F\left( R\right) \right| },\vec{0}\right) $
is the (timelike) unit vector normal to hypersurface $\Sigma $.

Integrating the charge density over the hypersurface $\Sigma\equiv S^n$, we
obtain the total electric charge at future/past infinity 
\begin{eqnarray}
\mathbf{Q}=-\frac{1}{4\pi }\oint_{\Sigma ^{\pm }}d^{n}\phi ^{\pm }\sqrt{%
\omega ^{\pm }}n^{\pm i}h_{i}^{\mu}F_{\mu \nu}^{\pm }u^{\pm
\nu}=2Q\omega_{n}^{-1}\sqrt{\frac{2(n-1)}{n}},
\end{eqnarray}
where $\omega_{n}=16\pi /(n\mbox{Area}(S^{n}))$.

The interpretation of this charge is analogous to that for the other
conserved quantities noted above. The charge $\mathbf{Q}$ is that measured
by a collection of observers traversing a given history; conservation of
charge implies that observers following a different history would measure
the same value of the charge.

The computation of a finite action for RNdS black holes is only a slight
generalization of the results presented in ref.~\cite{Ghezelbash:2002ab}.
Considering the EM equations of motion, we obtain the on shell bulk action 
\begin{eqnarray}
I_{B} &=&\frac{1}{16\pi }\int_{\mathcal{M}}d^{n+2}x~\sqrt{-g}\left( \frac{%
4\Lambda }{n}-\frac{2}{n}F^{2}\right) =  \nonumber  \label{actionbulk1} \\
&=&\frac{(n+1)H^{2}}{8\pi }\int d^{n+1}x\int_{R_{c}}^{R_{m}}dR\frac{\sqrt{h}%
}{\sqrt{\left| F(R)\right| }}-\frac{1}{8\pi n}\int
d^{n+1}x\int_{R_{c}}^{R_{m}}dR\frac{\sqrt{h}}{\sqrt{\left| F(R)\right| }}%
F^{2}.
\end{eqnarray}%
In this expression the integration is from the cosmological horizon out to
some $R_{m}$ that will eventually be sent to infinity. We shall work here in
the upper patch outside of the cosmological horizon in RNdS spacetime;
results for the lower patch are obtained in a similar manner. The second
term in the above relation is the usual electromagnetic field contribution.
In computing it we keep $A_{T}$ fixed at infinity, for a value of the
constant appearing in eq.~(\ref{A}) corresponding to the electrostatic
difference between the cosmological horizon and infinity 
\[
\Phi =-\frac{n\mathbf{Q}}{4(n-1)}\frac{\omega _{n}}{R_{c}^{n-1}}, 
\]%
which implies that the boundary term from the gauge field will vanish. A
general expression for the rest of the action has been derived in ref.~\cite%
{Ghezelbash:2002ab}, for a generic $F(R)$.

In the large $R$ limit we find the finite total action (after including the $%
I_{\partial B}$ and $I_{ct}$ contributions) 
\[
I=\frac{\beta \mbox{Area}(S^{n})}{8\pi }\left[ M+H^{2}R_{c}^{n+1}-\frac{n}{%
H^{n-1}}\ \frac{\Gamma (\frac{2p-1}{2})}{2\sqrt{\pi }\Gamma (p+1)}\delta
_{2p,n+1}\right] -\frac{1}{n}\Phi \mathbf{Q}, 
\]%
where $\beta $ is the periodicity of the coordinate $T$. This periodicity is
usually fixed by Wick rotating and requesting the absence of conical
singularities at $R=R_{c}$. The Wick rotation implies also a convergent path
integral and allows us to compute (\ref{Partitionaction}) in the
semi-classical approximation. Inserting these results in eqs. (\ref{W}) and (%
\ref{S}) we find a black hole entropy 
\[
~S=\frac{\beta \mbox{Area}(S^{n})}{8\pi }\Big(M(1-n)+H^{2}R_{c}^{n+1}\Big)-%
\frac{n-1}{n}\beta \Phi \mathbf{Q}=\frac{R_{c}^{n}\mbox{Area}(S^{n})}{4}, 
\]%
while the Hawking temperature of the cosmological horizon is 
\[
T_{c}=\frac{1}{\beta }=\frac{1}{2\pi R_{c}}\left( (1-n)\frac{M}{R_{c}^{(n-1)}%
}+H^{2}R_{c}^{2}-(1-n)\frac{Q^{2}}{R_{c}^{2(n-1)}}\right) . 
\]%
The same expression is obtained by equating it to the surface gravity on the
horizon. These quantities can be shown to be in accord with the first law of
thermodynamics 
\[
dE=TdS+\Phi d\mathbf{Q} 
\]
for the cosmological horizon.

Subtracting off the vacuum energy contribution for the Casimir energy (which
is non-zero only when $n$ is odd) we find 
\begin{equation}
\mathfrak{M}=-\frac{\mbox{Area}(S^{n})nM}{8\pi }=-\frac{R_{c}^{n-1}}{\omega
_{n}}\left( 1-H^{2}R_{c}^{2}+\frac{Q^{2}}{R_{c}^{2(n-1)}}\right) ,
\label{mass}
\end{equation}%
for the gravitational mass, measured at the far past or far future boundary
of dS space. Note that it is negative provided $M>0$, consistent with the
expectation \cite{Balasubramanian:2001nb} that pure dS spacetime has the
largest mass for a singularity-free spacetimes of this topology. If there is
a CFT dual to RNdS, this mass translates into the energy of the dual CFT
living on a Euclidean cylinder $R\times S^{n}$. The existence of a
macroscopic scale in the system explains the appearance of the Casimir term
(proportional with the central charge in two dimensions) in the expression
of the gravitational energy (\ref{energy}).

As in the AdS/CFT correspondence, the metric of the manifold on which the 
\textit{putative} dual CFT resides is defined by 
\[
\gamma _{ij}=\lim_{R\rightarrow \infty }\frac{1}{H^{2}R^{2}}h_{ij}. 
\]%
The dual field theory's stress tensor, $\tau _{k}^{i}$, is related to the
boundary stress tensor by the rescaling \cite{Myers:1999qn} 
\[
\sqrt{\gamma }\gamma ^{ij}\tau _{jk}=\lim_{R\rightarrow \infty }\sqrt{h}%
h^{ik}T_{jk}. 
\]%
The geometry of the manifold on which the dual CFT resides is given by the
cylinder metric 
\begin{eqnarray}
ds^{2}=\gamma _{ij}dx^{i}dx^{j}=dT^{2}+\frac{1}{H^{2}}d\Omega _{n}^{2}.
\end{eqnarray}
The CFT's stress tensor is 
\begin{eqnarray}
\tau _{j}^{i}=-\frac{1}{8\pi }(M-\frac{1}{H^{n-1}}\ \frac{\Gamma (\frac{2p-1%
}{2})}{2\sqrt{\pi }\Gamma (p+1)}\delta _{2p,n+1})(-(n+1)u^{i}u_{j}+\delta
_{j}^{i}),  \label{staticstress}
\end{eqnarray}
where $u^{i}=\delta _{T}^{i}$. This tensor is finite, covariantly conserved
and manifestly traceless. The trace anomaly indicates the quantum breaking
of conformal invariance in curved spaces and is proportional to both the
curvature and central charge in two-dimensions. The sphere $S^{n}$ has a
scale, $1/H^{2}$, but the conformal invariance of the theory is preserved
since this scale enters in a conformally invariant way in the metric itself %
\cite{awad}. Consequently, we obtain a traceless stress tensor (\ref%
{staticstress}).

Working outside the cosmological horizon, there is no manifestation of the
black hole horizon in the expression of the CFT's stress tensor. We will see
in Section 4 that, using planar coordinates which cover more of the bulk
manifold, the black holes will appear like punctures in the dual theory
corresponding to divergences in the dual stress tensor.

Verlinde \cite{verlinde} generalized the Cardy formula, that counts the
quantum states of a two-dimensional CFT, to any dimension. This result, now
commonly referred to as the Cardy--Verlinde formula, is supposed to be a
formula yielding the entropy for a CFT in any dimension. As discussed in
ref. \cite{Cai:2001tv} (see also refs. \cite{{halyo},{medved}}), the entropy
of the cosmological horizon of a RNdS solution can be rewritten in the form
of the Cardy--Verlinde formula: 
\begin{eqnarray}
S=\frac{2\pi }{nH}\sqrt{|E_{c}|(2(\mathfrak{M}-E_{q})-E_{c})},  \label{ent1}
\end{eqnarray}
where $E_q$ is the energy of the $U(1)$ field 
\[
E_{q}=\frac{1}{2}\Phi \mathbf{Q}=-\frac{Q^{2}}{\omega _{n}R_{c}^{n-1}} 
\]%
and the Casimir energy $E_{c}$, defined in this context as the violation of
the Euler identity $E_{c}=(n+1)\mathfrak{M}-nTS-n\Phi \mathbf{Q}$, is 
\begin{equation}
E_{c}=-\frac{2nR_{c}^{n-1}\mbox{Area}(S^{n})}{16\pi }.  \label{casimir}
\end{equation}%
These results are usually interpreted as providing support for the dS/CFT
correspondence. Note that, unlike the AdS/CFT case, the Casimir energy is
negative, indicating the non-unitarity of the dual CFT \cite%
{Strominger:2001pn}.

Similar thermodynamic quantities can be defined also for the black hole
horizon, while the total entropy of the RNdS spacetime is believed to be the
sum of entropies of the black hole and cosmological horizons. Since the
total entropy of the RNdS spacetime is smaller than that of empty dS
spacetime, the black hole is made by borrowing degrees of freedom from the
horizon of empty dS spacetime and freezing them into special configurations.

As observed in ref. \cite{Cai:2001tv}, if one uses the BBM definition of the
mass, the black hole entropy cannot be rewritten in a Cardy--Verlinde form.
However, when the Abbot--Deser prescription for the total mass is used
(which gives a positive gravitational mass), the black hole entropy can also
be written in a form similar to eq.~(\ref{ent1}).

We conclude that the CFT on $\mathcal{I}^{+}$ accounts for the entropy of
the cosmological horizon only, in accord with the fact that the
thermodynamic quantities are measured by observers outside the cosmological
horizon.


\subsection{Planar coordinates}

Although the line element (\ref{rnds}) takes a simple form in a static
coordinate system, the expansion of the universe is not manifest.
Furthermore, the static coordinates $(T,R)$ break down at the Killing
horizons. However RNdS black holes admit a simple expression in planar
coordinates 
\begin{eqnarray}
&&ds^{2}=a^{2}U^{\frac{2}{n-1}}(dr^{2}+r^{2}d\Omega _{n}^{2})- \frac{V^{2}}{%
U^{2}}dt^{2},  \label{metric} \\
&&a=e^{Ht},~~V=1-\frac{\alpha }{\rho ^{2}},~~U=1+\frac{M}{\rho }+\frac{
\alpha }{\rho ^{2}},  \nonumber
\end{eqnarray}
where $\rho =(ar)^{n-1}$ and $\alpha =(M^{2}-Q^{2})/4$.

For $n>2$, the metric (\ref{metric}) generalizes the four dimensional
solutions discussed in ref. \cite{Brill:1993tw}. For electrically charged
black holes, the only nonvanishing component of the $U(1)$ potential is: 
\begin{eqnarray}
A_{t}=Q\sqrt{n(n-1)/2}\int a^{1-n}V(U^{2}r^{n})^{-1}dr.
\end{eqnarray}
Using planar coordinates has the disadvantage that the manifest
time-translation symmetry is broken and the charts that cover the horizons
are highly distorted. The coordinate transformation relating the static and
planar metrics is 
\begin{eqnarray}
R=arU^{\frac{1}{n-1}},~~t=T+h(R),~~\mathrm{with}~~h^{\prime }(R)=-\frac{HR}{%
F(R)\sqrt{F(R)+H^{2}R^{2}}}\ .  \label{transform}
\end{eqnarray}
The time-dependence enters in a natural manner in this solution and for
large $r$, the dS metric in planar coordinates is approached. We note that
the case $Q=0$ is just the McVittie solution describing a Schwarzschild
black hole embedded in $(n+2)$--dimensional de Sitter spacetime \cite%
{Shiromizu:2001bg}. A discussion of the global structure of the line element
(\ref{metric}), as well as the the coordinate transformation (\ref{transform}%
) appears in ref. \cite{Brill:1993tw} for $n=2$. It is straightforward to
show that these general features remain valid for $n>2$, where the analysis
depends crucially on the sign of $H$. Analogues with pure dS spacetime (see
Fig. 1(a)), cosmologically expanding ($H>0$) and contracting ($H<0$)
coordinate systems also exist in RNdS spacetime, being centered only about
the black hole. Finally we note that when $M>M_N$ (the ``overmassive'' case)
the global structure is strongly modified as in Fig. 1(b), and a timelike
naked singularity appears.

\begin{center}
\includegraphics[scale=0.8]{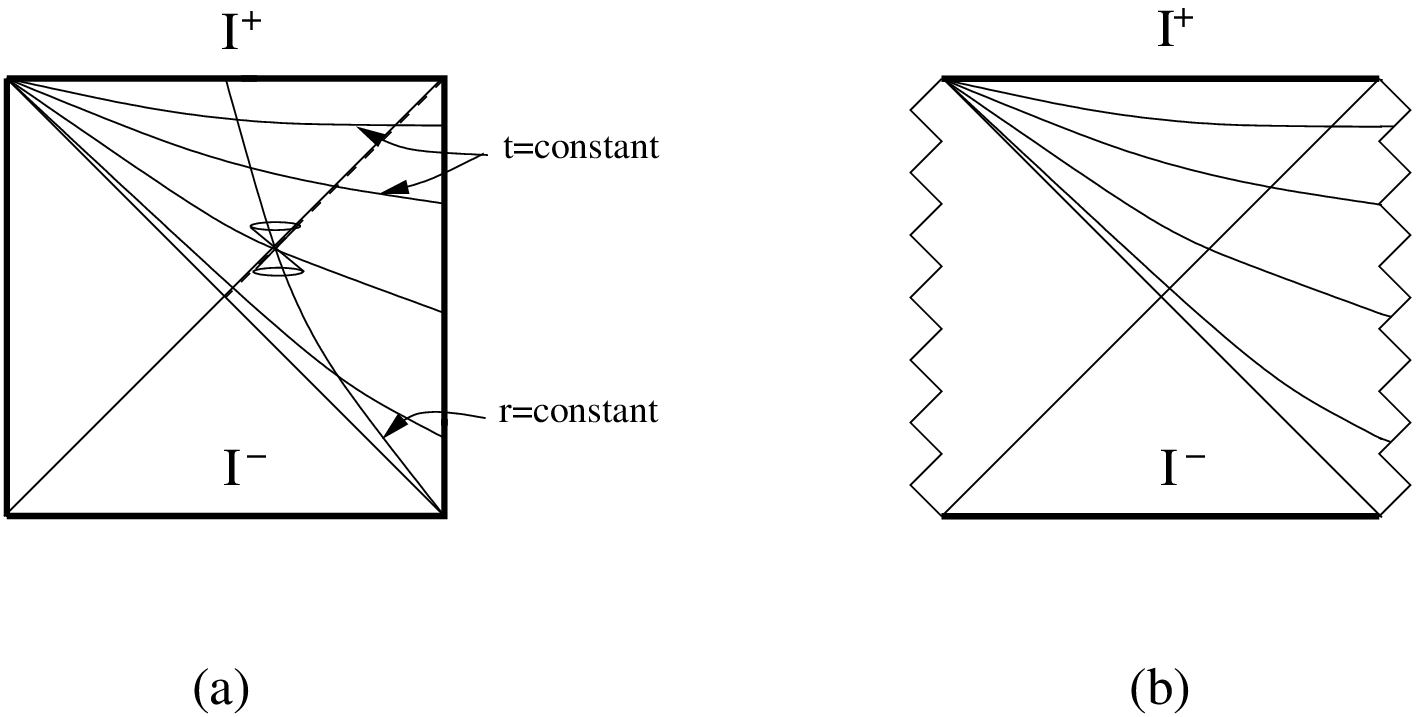}
\end{center}

Figure 1: Carter--Penrose diagram for pure dS spacetime (a) and for the
``overmassive'' case (b). Singularities are represented by wavy lines.

\bigskip

Although the singularity is naked, this is not a violation of cosmic
censorship. The reason is that we start with bad singular initial conditions
that imply the naked singularity exists at any time. The outer black hole
and the cosmological horizons have disappeared. However the inner black hole
horizon can be interpreted as a cosmological one, separating the two naked
singularities at antipodal regions of a background dS spacetime.

The geometry (\ref{metric}) is preserved by the transformation 
\begin{eqnarray}  \label{gen}
t \to t +\lambda,~~ r \to e^{-\lambda H} r.
\end{eqnarray}
In the above relation, the first term generates time evolution in the bulk
gravity theory, while the second term generates scale transformations in the
boundary theory. The Killing vector associated with this symmetry reads 
\begin{eqnarray}  \label{KV}
\xi=-Hr\frac{\partial}{\partial r}+\frac{\partial}{\partial t}
\end{eqnarray}
and is not globally timelike. $\mathcal{I}^{\pm }$ are now approached for
large $\left| t\right| $, while the boundary topology is $R^{n+1}$. 
For $H>0$, we take the Euclidean CFT to live at future infinity $\mathcal{I}%
^{+}$ on the space parametrized by $x^{i}$. From the UV/IR correspondence,
an object at time $t$ in dS space corresponds to an excitation of scale size 
$\delta x=e^{Ht}$ in the CFT. The CFT resides on the Euclidean line element 
\begin{eqnarray}
ds^{2}=\gamma _{ij}dx^{i}dx^{j}=dr^{2}+r^{2}d\Omega _{n}^{2}.
\end{eqnarray}
A straightforward calculation following the BBM prescription gives the
boundary stress-tensor associated with the line-element (\ref{metric}): 
\begin{eqnarray}
T_{rr} &=&-\frac{1}{8\pi H}\frac{n}{n-1}\frac{U_{,r}}{rU}\left( 1+\frac{r}{%
2(n-1)}\frac{U_{,r}}{U}\right) =\frac{M}{8\pi Hr^{2}}\frac{n}{(ar)^{n-1}}+O(%
\frac{1}{a^{n}}),  \label{tik1} \\
T_{ab} &=&-\frac{r^{2}\omega _{ab}}{8\pi H}\frac{1}{n-1}\left( -\frac{n}{n-1}%
\Big(\frac{U_{,r}}{U}\Big)^{2}+(n-1)\frac{U_{,r}}{rU}+\frac{U_{,rr}}{U}%
\right) =-\frac{M}{8\pi H}\frac{\omega _{ab}}{(ar)^{n-1}}+O(\frac{1}{a^{n}}).
\nonumber
\end{eqnarray}%
We arrived at the above result using the fact that for the planar slicing
the extrinsic curvature is proportional with the intrinsic metric for each
slice, $K_{ij}=-Hh_{ij}$, and the first three terms in the boundary stress
tensor exactly cancel.

The stress tensor of the dual field theory has the simple form 
\begin{eqnarray}
\tau _{r}^{r}=\frac{M}{8\pi H}\frac{n}{r^{n+1}},~~\tau _{b}^{a}=-\delta
_{a}^{b}\frac{M}{8\pi H}\frac{1}{r^{n+1}}.  \label{planarstress}
\end{eqnarray}

This tensor is covariantly conserved and manifestly traceless, as expected
for a conformal field theory on a manifold with zero curvature. The presence
of the divergent factor $1/r$ in the above expression reflects the
inhomogeneity of the bulk geometry.

The quantity we need is the mass of the black hole as an excitation in $%
dS_{n+2}$. The mass is the charge associated with the Killing vector (\ref%
{KV}), measured at the far past or far future (depending on the sign of $H$%
): 
\[
\mathfrak{U}=\mathfrak{M}=-\frac{\mbox{Area}(S^{n})nM}{8\pi }, 
\]%
yielding a value for the energy of the dual CFT which agrees with (\ref{mass}%
). The negative sign implies that the black hole lowers the total bulk
energy with respect to the total energy of the pure dS space, which is zero
in planar coordinates. The absence of a macroscopic scale for a boundary
with topology $R^{n+1}$, unlike the static-coordinate case when the boundary
topology is $R\times S^{n}$, leads to the absence of a Casimir contribution %
\cite{Shiromizu2002}.

The norm of the Killing vector vanishes at the horizons, which are again the
dS horizon $r_{c}$ and the outer black hole horizon $r_{+}$, being located
at the solutions of the equation $\xi _{k}\xi ^{k}=0$. However, by using the
coordinate transformation (\ref{transform}) we find that $\xi _{k}\xi ^{k}=0$
corresponds to $F(R)=0$, which is the equation for the position of a horizon
in static coordinates. Thus, although the location of a horizon is
time-dependent in planar coordinates, the product $r_{h}a$ is constant. The
event horizon radius in static coordinates $R_{h}$ is related to $\rho
_{h}=(ar_{h})^{n-1}$ through the relation $M+\rho _{h}+\alpha /\rho
_{h}=R_{h}^{n-1}$.

The thermodynamic properties of the RNdS black hole in planar coordinates
can also be discussed, at least formally. The concept of assigning a
temperature to dS space is well-defined only in the static patches. However,
we can assign a Hawking temperature to these event horizons by using the
standard relation $T=|\kappa |/2\pi ,$ where $\kappa $ is the surface
gravity of the horizons, computed from 
\[
\frac{1}{2}\nabla _{k}(\xi _{i}\xi ^{i})=-\kappa \xi _{k}. 
\]%
The values of the Hawking temperature for both the black hole and
cosmological horizons are the same in both coordinate systems. We can prove
that by using the properties of the coordinate transformation (\ref%
{transform}) 
\[
\frac{dF(r(R,T))}{dr}\bigg|_{r=r_{c}}=\frac{dF(R)}{dR}\frac{dR}{dr}\bigg|%
_{r=r_{c}},~~~~\frac{dR}{dr}=\frac{VR}{Ur}. 
\]%
Furthermore, the event horizon area is the same in both coordinate systems
since $A=(r_{h}aU^{\frac{1}{n-1}})^{n}\mbox{Area}(S^{n})$, and from eq.~(\ref%
{transform}), $A=R_{h}^{n}\mbox{Area}(S^{n})$.

Thus, we are motivated to associate an entropy $S=A/4$ with the horizons of
the cosmological RNdS solution (\ref{metric}). Following Bousso's
formulation of the holographic principle \cite{Bousso:2000md}, an entropy
bound for empty dS space in planar coordinates has been discussed in ref.~%
\cite{Kabat:2002hj}. However, similar arguments can be used here also. The
entropy contained at time $t$ in a spatial ball of arbitrary radius $%
x^{i}x^{i}\leq \tilde{r}^{2}$ is bounded by \cite{Bousso:2000md}, 
\[
S=\frac{\mathrm{surface~area~of~ball}}{4}=\frac{\mbox{Area}(S^{n})}{4} (a 
\tilde{r} U^{\frac{1}{n-1}})^{n}. 
\]%
For $\tilde{r}=e^{-Ht}U^{\frac{1}{1-n}} R_c $, $i.e.$ when the surface ball
coincides with the cosmological horizon, this is exactly the entropy of the
RNdS space in static coordinates.

Thus, although the horizon and entropy in dS space have an obvious observer
dependence, thermodynamic quantities defined in this way are the same as
those in the case of static coordinates.

We can further demonstrate that the dS horizon entropy adopts the
Cardy--Verlinde-like form (\ref{ent1}). In deriving the expression of the
chemical potential $\Phi $ we make use of the gauge properties of the $U(1)$
field, and similar to the static case we express all the thermodynamic
quantities in terms of $(Q,~R_{c},~H)$. Therefore we find the same general
picture for two different patches, with different classes of observers. We
can interpret this result as providing support for the dS/CFT
correspondence, since the general features of the CFT dual to a RNdS black
hole should not depend on the dS slicing choice.


\section{Multi-black hole solutions}

A case of particular interest is $Q^{2}=M^{2}$. As originally found by
Kastor and Traschen in four dimensions \cite{Kastor:1992nn}, this extremal
RNdS metric admits a generalization to $N$ black holes, with the
line-element \cite{London:ib} 
\begin{eqnarray}
ds^{2} &=&a^{2}\Omega ^{\frac{2}{n-1}}\big((dx^{1})^{2}+...+(dx^{n+1})^{2}%
\big)-\Omega ^{-2}dt^{2},  \label{Nbh} \\
\Omega &=&1+\frac{1}{a^{n-1}}\sum_{A=1}^{N}\frac{M_{A}}{|r-r_{A}|^{n-1}}%
,~~a(t)=e^{Ht}.  \nonumber
\end{eqnarray}%
Here $|r-r_{A}|$ denotes the Euclidean distance between the field point $r$
and the fixed location $r_{A}$ in a Euclidean space of cosmological
coordinates: $|r-r_{A}|=(\sum_{k=1}^{n+1}(x^{k}-x_{A}^{k})^{2})^{1/2}$. The
black hole locations, $r_{A}$, are arbitrary, while the Maxwell one-form is
given by $A=A_tdt=\Omega \sqrt{n/(2(n-1))}dt$.

The KT solution is time dependent and has no symmetries in general. A
generalization of this solution to include a dilaton was given in ref. \cite%
{Maki:1992tq}.

In the limit of vanishing cosmological constant, the KT solution reduces to
the well-known static Majumdar--Papapetrou solution \cite{hah}. In contrast,
the black holes with $\Lambda >0$ are highly dynamical: they ignore one
another and follow natural trajectories in the background of dS space. For $%
H<0$, the KT solution describes a system of ``incoming'' charged black holes
(all with the same sign of charge), which collide and coalesce for a certain
range of parameters, providing a new arena in which to test cosmic
censorship. The solutions describing the time-reversed situation (a system
of ``outgoing'' white holes, oppositely charged) is obtained by changing the
sign of $H$ \cite{Kastor:1992nn,Brill:1993tm}. The ``incoming'' set is
described by cosmological coordinates that include $\mathcal{I}^{-}$,
whereas for the ``outgoing'' set, $\mathcal{I}^{+}$ is included.

The surfaces of constant $t$ are spacelike everywhere and the boundary
metric, approached for large $t$, is $ds^{2}=a^{2}\Omega ^{\frac{2}{n-1}}%
\big((dx^{1})^{2}+...+(dx^{n+1})^{2}\big)$. However, a multi-black hole
solution has no isometry of the form (\ref{gen}) and there is no quasilocal
conserved charge associated with $M$, nor any well-defined quasilocal
thermodynamic quantities.

A computation of the boundary stress tensor is still possible, with the
result 
\[
T_{ij}=-\frac{1}{8\pi H}\frac{1}{n-1}\left( \frac{n}{n-1}\frac{\Omega
_{,i}\Omega _{,j}}{\Omega ^{2}}-\frac{\Omega _{,ij}}{\Omega }+\delta _{ij}%
\Big(\frac{n-4}{2}\frac{\Omega _{,k}\Omega _{,k}}{\Omega ^{2}}+\frac{\Omega
_{,kk}}{\Omega }\Big)\right) . 
\]%
The corresponding stress tensor of the \textit{presumed} dual field theory
is covariantly conserved, traceless and has the form 
\begin{eqnarray}
\tau ^{ij}=-\frac{1}{8\pi H}\sum_{A=1}^{N}\frac{M_{A}}{|r-r_{A}|^{n+1}}%
\left( \delta ^{ij}-(n+1)\frac{(x^{i}-x_{A}^{i})(x^{j}-x_{A}^{j})}{%
|r-r_{A}|^{2}}\right) ,  \label{multistress}
\end{eqnarray}
diverging again at the black hole locations. Since the isometry (\ref{gen})
is recovered for $r\gg r_{A}$, we can define a mass only asymptotically.
Thus, in the limit of large $r$, and for $r_{A}$ in a compact region of
Euclidean coordinate space, the $N-$black hole solution approaches the
solution (\ref{metric}) with the mass parameter $M=\sum M_{A}$ and the total
gravitational mass is the sum of the individual masses 
\[
\mathfrak{M}=-\frac{\mbox{Area}(S^{n})n}{8\pi }\sum M_{A}=\sum \mathfrak{M}%
_{A}. 
\]%
Another case of interest is when $|r-r_{A}|$ is small, for fixed $A$. Let us
choose the origin so that $r_{A}=0$. The conformal factor of the boundary
metric 
\[
a(t)^{2}\Omega ^{\frac{2}{n-1}}\approx \left( a(t)^{n-1}+\frac{M_{A}}{%
|r|^{n-1}}+\sum_{B\neq A}\frac{M_{B}}{|r_{B}|^{n-1}}\right) ^{\frac{2}{n-1}%
}=\left( a(t)^{n-1}+\frac{M_{A}}{|r|^{n-1}}+\mathrm{{const}.}\right) ^{\frac{%
2}{n-1}} 
\]%
differs by a constant from the conformal factor of a metric that describes
just a black hole with mass parameter $M_{A}$. It is clear that an observer
on a boundary near to the $A^{th}$ black hole, will measure a gravitational
mass $\mathfrak{M}=\mathfrak{M}_{A}$ if the other black holes are far enough
away ($i.e.$ if $r_{B}$ is sufficiently large).

Unlike (\ref{staticstress}), the CFT's stress tensor in planar coordinates
--- see eqs. (\ref{planarstress}) and (\ref{multistress}) --- has
divergences. Consider the meaning\footnote{%
Here we recapitulate the discussion from ref.~\cite{Brill:1993tm} (see,
also, ref.~\cite{brill}), in the light of the dS/CFT correspondence.} of
these divergences in the $Q=M$ case for $n=2$; these interpretations are
also valid for general $n$. An analysis of these configurations similar to
the single-black hole case appears difficult --- their properties are better
understood by using a coordinate system with $\tau =e^{Ht}/H$ and continuing 
$\tau $ to negative values (leaving unchanged the form of the CFT stress
tensor (\ref{multistress})). The metric becomes \cite%
{Brill:1993tw,Brill:1993tm}: 
\[
ds^{2}=-\frac{d\tau ^{2}}{U(\tau ,r)^{2}}+U(\tau ,r)^{2}dr\cdot dr,~~~U(\tau
,r)=H\tau +\sum_{A}\frac{M_{A}}{|r-r_{A}|}. 
\]%
Let us work in expanding patch ($H>0$), that covers the expanding
cosmological region and part of the white hole. This is shown explicitly in
Figure 2(a). The black hole is ``undermassive'' ($M<M_{N}$), so all three
horizons (inner and outer black hole horizons and cosmological horizon) are
present. The boundaries of the chart are given by: the black hole event
horizon at $(\tau =+\infty ,r=r_{A})$, the inner horizon at $(\tau =-\infty
,r=r_{A})$, the singularity for $U=0$ and the $\mathcal{I}^{+}$ at ($\tau
=+\infty ,r=$ finite). The maximal analytic extension \cite{Brill:1993tw},
for one black hole, can be obtained by gluing an expanding and a contracting
cosmological patch at the cosmological horizon $(\tau =0,r=+\infty )$, as
illustrated in Figure 2(b).

\begin{center}
\includegraphics[scale=0.8]{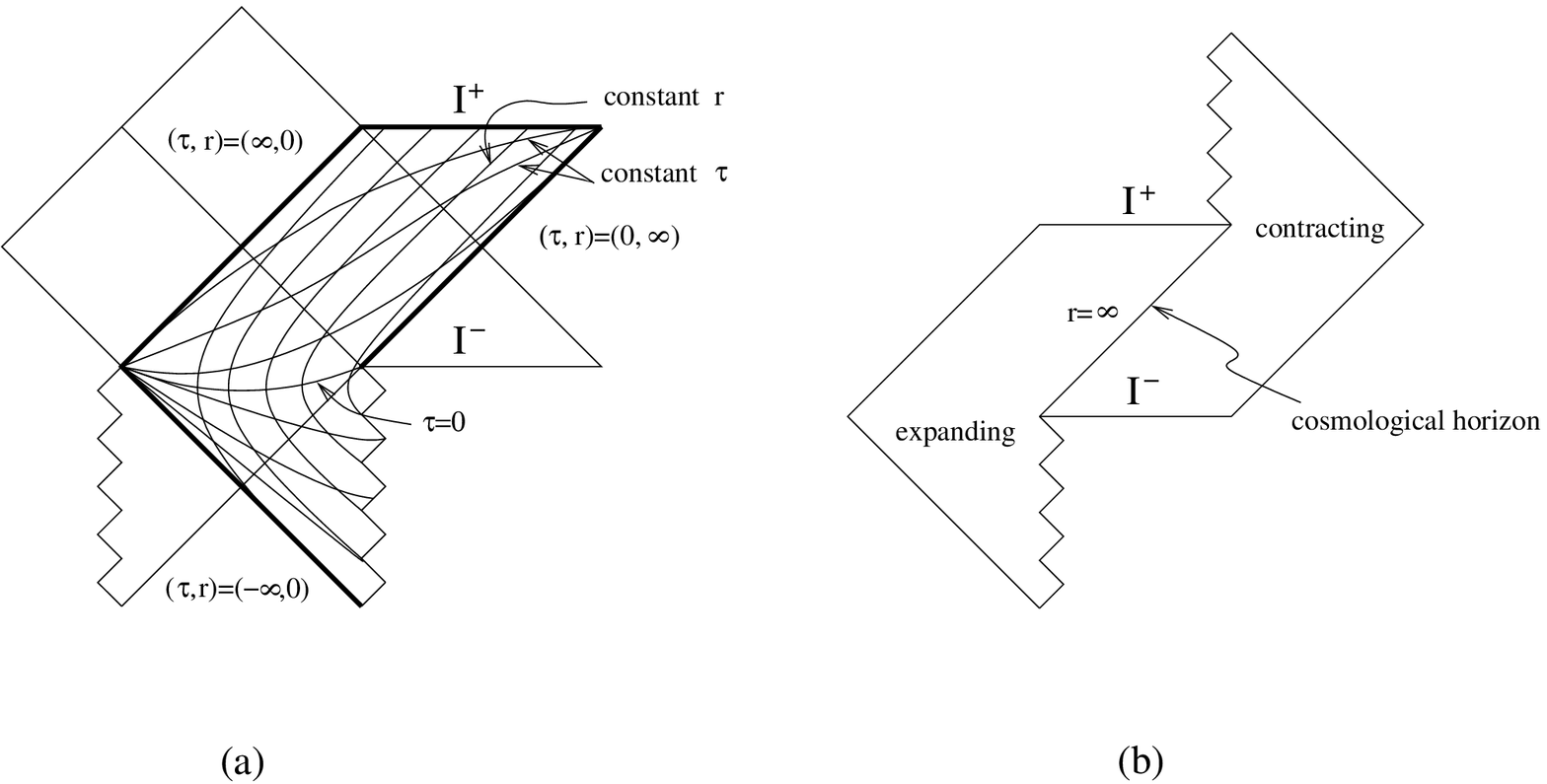}
\end{center}

Figure 2: Carter-Penrose diagram for an undermassive RNdS black hole in the
case Q=M at the location $r_A=0$. The region covered by the expanding chart
is shown in (a), the thick lines representing the boundaries of the chart
and the wavy lines singularities. An expanding and a contracting patch are
shown in (b), but the figure can be repetead indefinitely in the horizontal
and vertical direction, yielding a spacetime that is spacelike and timelike
periodic.

\bigskip

Near $r_A$ the metric describes a cylindrical geometry resembling the
infinite throat of the asymptotically flat RN. The slice $\tau=0$ is regular
and has a cylindrical form everywhere. In fact, we see that a singularity
appears at $r=+\infty$ (corresponding to $U(0,r=+\infty)=0$). The
singularity cuts off more and more of the cylinder, as $\tau$ is decreasing
to negative values, and at $\tau=-\infty$ the singularity surrounds the
throat at $r=r_A$. The slices given by $\tau>0$ are non-singular since $%
U(\tau,r)>0$ in this case. It is clear now that $\mathcal{I}^{+}$ is an
asymptotically flat Euclidean surface, with the cylindrical form of an
infinite throat near $r=r_A$. The dual CFT is defined on $\mathcal{I}^{+}$,
that is a plane with the ``point'' $r=r_A$ removed, and the black hole
appears like a puncture in the dual theory.

We can extend this discussion to the KT multi black-hole solution case. We
showed that in the limit of large $r$ the multi-black hole solution
resembles the RNdS solution with the gravitational mass given by the sum of
individual masses. We can still glue expanding and contracting patches at $%
r=+\infty $, but this can be done with varying degrees of smoothness. This
means that there is not a unique analytic extension\footnote{%
In the final section we comment on the physical interpretation of this lack
of smoothness and on cosmic censorship.}. $\mathcal{I}^{+}$ is again a
non-singular and asymptotically flat at $r=+\infty $ with the form of an
infinite throat around each ``point'' $r_{A}$, as in Figure 3.

\begin{center}
\includegraphics[scale=0.3]{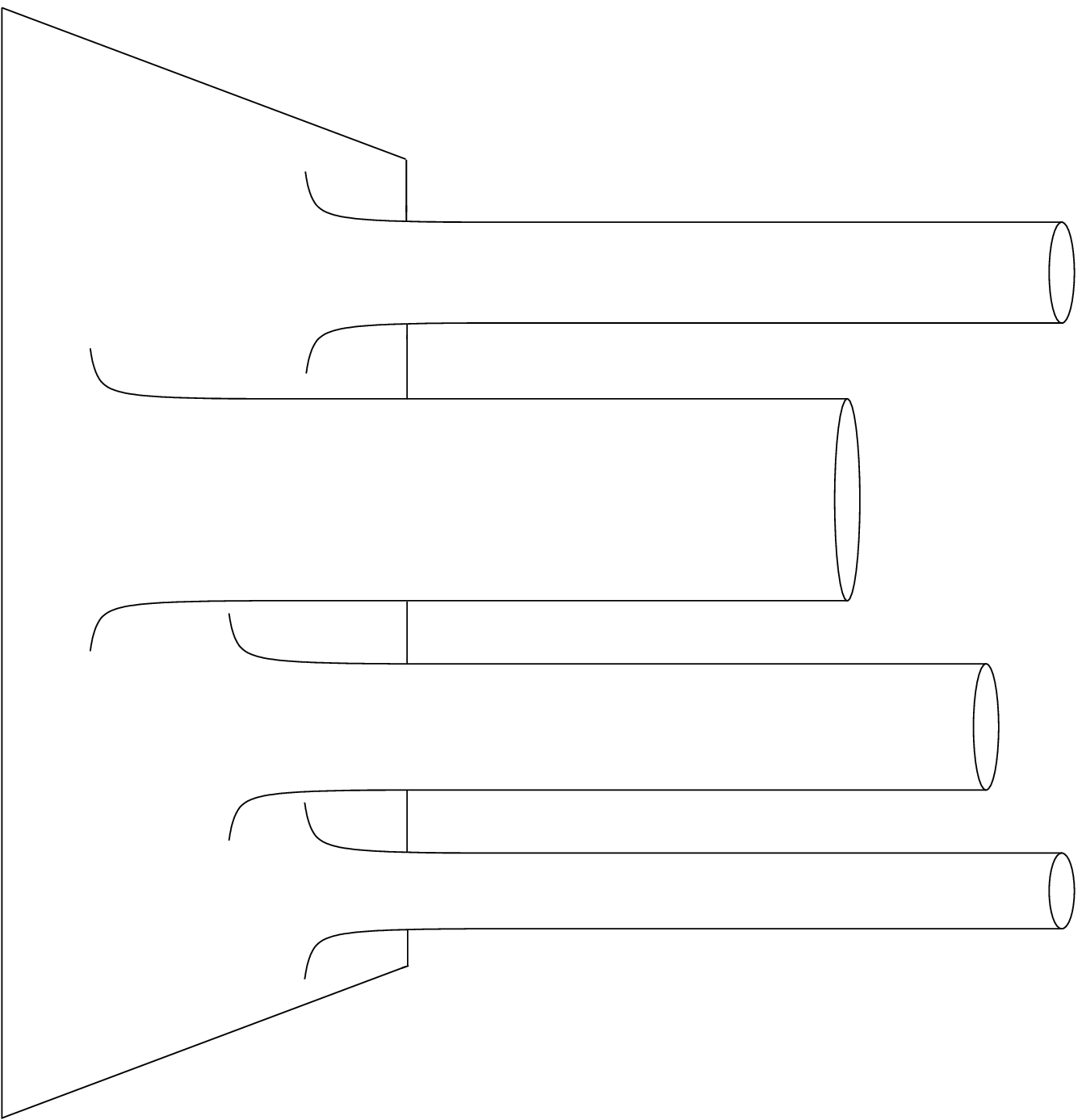}
\end{center}

Figure 3: The surface $\mathcal{I}^{+}$ of an expanding KT spacetime. The
missing ``points'' represent disjoint boundary components. In the expanding
patch the black holes remain separately for all times.

\bigskip

The interpretation of the divergences in the stress tensor of the CFT is
that the black holes appear as punctures in the dual theory.

\section{RG flow and $c$-function}

A remarkable property of the AdS/CFT correspondence is that it works even
far from the conformal regime: supergravity describes an $N=1$
supersymmetric renormalization group (RG) flow and there is a $c$-function
that is monotonic decreasing from the ultraviolet (UV) regime (large radii
in AdS space) to the infrared (IR) regime (small radii in the AdS space) of
the dual CFT (see, $e.g.$, refs.~\cite{kraus, gubser,odint}). This result is
consistent with the interpretation of the radial coordinate of AdS space as
a energy scale: the CFT on the boundary is scale invariant and the QFT has a
scale (energy) $\mu $. The RG ``trajectory'' allows us to define the UV
limit ($\mu \rightarrow \infty $) and the IR limit ($\mu \rightarrow 0$) of
a given QFT. In the AdS/CFT context, the CFT on the boundary is the UV fixed
point of a QFT in the bulk. The central charge counts the number of massless
degrees of freedom in the CFT (it counts the ways in which energy can be
transmitted). The coarse graining of a quantum field theory removes the
information about small scales, in other words there is a gradual loss of
non-scale invariant degrees of freedom.

In ref.~\cite{kraus}, the interior holographic duals for AdS were
investigated by adopting a Wilsonian RG perspective. That is, to foliate AdS
space with surfaces of constant radial coordinate, choose a slice that
enclosed a specific volume and perform the bulk path integral over the
excluded volume. In this way, a subset of the system is described by
``integrating out'' the excluded degrees of freedom. Different foliations of
the spacetime lead to different RG flows and in general there is no
requirement that a field should be coarsened uniformly.

In what follows\footnote{%
We thank Rob Myers for suggestions and discussions on this section.}, we use
these ideas to understand how the RG flow is modified when a charged black
hole exists in dS space and also how it is changed by different foliations%
\footnote{%
We note that holographic RG flows have been used in the context of the
dS/CFT correspondence to calculate the holographic conformal anomaly \cite%
{nojodint}.}. In analogy with the AdS/CFT case, we can interpret the time
evolution in asymptotically dS spaces as a RG flow \cite%
{Balasubramanian:2001nb,strominger}. A step further was made in ref.~\cite%
{fred}, where a $c$-function, which can be evaluated on each slice of some
foliation, was defined. The idea is to associate an effective cosmological
constant 
\[
\Lambda _{eff}=G_{\mu \nu }n^{\mu }n^{\nu }=\rho 
\]%
to any slice that can be embedded in inflationary dS space; here $n^{\mu }$
is the unit normal vector to a constant time slice and $\rho $ is the energy
density on a constant time hypersurface. By dimensional analysis and
imposing the $c$-function to be a function of this effective cosmological
constant, one finds: 
\begin{equation}
c\sim \frac{1}{\Lambda _{eff}^{n/2}}=\rho ^{-\frac{n}{2}}.  \label{cfn}
\end{equation}%
In ref.~\cite{thorlacius}, the c-function is geometrically interpreted as
the area of the apparent cosmological horizon in Planck units. The
generalized dS $c$-theorem states that in a contracting patch of dS
spacetime, the RG flows toward the infrared and in an expanding spacetime,
it flows toward the ultraviolet.

To compute $\Lambda _{eff}$ we make use of the Einstein equations 
\begin{eqnarray}
G_{\mu }^{\nu }+\Lambda \delta _{\mu }^{\nu } =2T_{\mu }^{\nu },  \nonumber
\end{eqnarray}
where $T_{\mu }^{\nu } =F_{\mu \alpha }F^{\alpha \nu }-\frac{1}{4}%
F^{2}\delta _{\mu }^{\nu }$ is the energy-momentum tensor of the gauge
field. In static coordinates, the only nonzero component of the gauge field
is $A_{T}$, and so 
\[
F^{2}=2F_{RT}F^{RT}=-\frac{n^{2}\omega _{n}}{8R^{2n}}\mathbf{Q}^{2}. 
\]%
However, this is an invariant, valid also in planar coordinates and we
obtain $T_{t}^{t}=T_{T}^{T}=\frac{1}{4}F^{2}$. Thus, the effective
cosmological constant in static coordinates is given by 
\begin{equation}
\Lambda _{eff}=-G_{T}^{T}=\Lambda -2T_{T}^{T}=\Lambda -\frac{1}{2}%
F^{2}=\Lambda +\frac{n^{2}\omega _{n}}{16R^{2n}}\mathbf{Q}^{2}
\label{Lameff}
\end{equation}%
and is larger in the bulk than on the boundary ($R\rightarrow +\infty $).
The corresponding $c$-function is decreasing from the (conformal) boundary
to the bulk and has finite values at $\mathcal{I}^{\pm }$ and at the
cosmological horizon \cite{fred}. For $Q=0$, the Schwarzschild-dS vacuum
solution is obtained and the RG flow is trivial corresponding to a constant $%
\Lambda _{eff}=\Lambda $.

In planar coordinates $(r,t)$, the effective cosmological constant is given
by 
\begin{eqnarray}
\Lambda_{eff}=-G_t^t=\Lambda-2T_t^t=\Lambda-\frac{1}{2}F^2.
\end{eqnarray}
It is easy to demonstrate that 
\begin{eqnarray}
\frac{\partial T_t^t}{\partial t}= 4n(n-1)^2\mathbf{Q}^2\frac{1}{R^{2n+1}}%
\frac{\partial R}{\partial t}~,~~~\frac{\partial R}{\partial t}=HVR/U.
\end{eqnarray}

In the case $Q=M$, the sign of $\frac{\partial T_{t}^{t}}{\partial t}$ is
given by the sign of $H$. Therefore, in a contracting (expanding) patch $H<0$
($H>0$) the effective cosmological constant is a monotonically increasing
(decreasing) function of $t$ and the corresponding $c$-function is a
monotonically decreasing (increasing) function of $t$. As we expected, the
RG flow in planar coordinates is not uniform because the induced metric on a
slice $t=const.$ has also an $r$ dependence. This resembles the lattice
field theory case where meshes of different sizes are used in different
regions \cite{kraus}.

\section{From dS to AdS space}

In static coordinates, we can formally ``Wick rotate'' from a positive to a
negative cosmological constant by using the analytic continuation $%
H\rightarrow -iH$. Together with $T\rightarrow iT$ this generates a solution
in the Euclidean AdS space (note also that, for a real EM solution we must
analytically continue the electric charge $Q\rightarrow iq$). However there
are different analytic continuations that generate solutions with Lorentzian
signatures. As discussed in ref.~\cite{Balasubramanian:2002am} in the
AdS/CFT context, the continuation $\theta \rightarrow \pi /2+i\tau $ gives a
bubble spacetime with a line element 
\begin{eqnarray}
ds^{2} &=&\frac{dR^{2}}{F(R)}+R^{2}\left( -d\tau ^{2}+\cosh ^{2}\tau d\Omega
_{n-1}^{2}\right) +F(R)dT^{2},  \label{b1} \\
F(R) &=&1-\frac{2M}{R^{n-1}}-\frac{q^{2}}{R^{2(n-1)}}+H^{2}R^{2}.  \nonumber
\end{eqnarray}%
Bubble spacetimes have previously been considered by various authors as
examples of time-dependent backgrounds with negative cosmological constant %
\cite{Balasubramanian:2002am}-\cite{Aharony:2002cx}. Typically, these
spacetimes are obtained by double analytic continuations of asymptotically
AdS (or flat) black holes. 

However, in special cases, similar constructions can be obtained starting
with dS solutions \cite{Birmingham:2002st}. For the RNdS solution in
cosmological coordinates (\ref{metric}), the transformation $t \to i
\chi,~~H \to -i H,~~Q \to iq$ generates a Euclidean solution of the EM
equations with negative cosmological constant. For $M=Q=0$ this gives the
Poincare patch of $AdS_{n+2}$.

A simple analytic continuation yielding a Lorentzian line-element is again $%
\theta \rightarrow \pi /2+i\tau $. The resulting spacetime is 
\begin{eqnarray}
ds^{2} &=&\frac{V^{2}}{U^{2}}d\chi ^{2}+a^{2}U^{\frac{2}{n-1}%
}(dr^{2}+r^{2}\left( -d\tau ^{2}+\cosh ^{2}\tau d\Omega _{n-1}^{2})\right)
\label{ads1} \\
V &=&1-\frac{M^{2}+q^{2}}{4(ar)^{2(n-1)}},~~U=1+\frac{M}{(ar)^{n-1}}+\frac{%
M^{2}+q^{2}}{4(ar)^{2(n-1)}},~~~a=e^{H\chi },  \nonumber
\end{eqnarray}%
which is a solution of the EM equations with negative cosmological constant.
The coordinate transformation between the known RN-AdS bubble solution (\ref%
{b1}) and the line element (\ref{ads1}) reads 
\[
R=arU^{\frac{1}{n-1}},~~t=T+h(R),~~\mathrm{with}~~h^{\prime }(R)=-\frac{HR}{%
F(R)\sqrt{F(R)-H^{2}R^{2}}}. 
\]%
The spacetime (\ref{ads1}) describes a ``bubble'' in an asymptotically
anti-de Sitter spacetime, which contracts from infinite size at $\tau
\rightarrow -\infty $ to a minumum size as $\tau =0$ and then expands back
out to infinite size as $\tau \rightarrow \infty $. One can prove that the
metric (\ref{ads1}) is smooth as $(ar)^{2(n-1)}\rightarrow \alpha $ (or $%
V\rightarrow 0$). Consequently there is no intrinsic periodicity of the
coordinate $\chi $; indeed there cannot be or else the metric will not be
single-valued. We can understand this by observing that the region around $%
R_{h}$ (with $F(R_{h})=0$) in eq.~(\ref{b1}) is not covered by the
coordinate system ($r,\chi $) (since $V^{2}/U^{2}=F(R)-H^{2}R^{2}$). Since $%
\chi $ is not periodic, the bubble does not close back on itself --- in this
sense it is not really a bubble. However the metric (\ref{ads1}) should be a
valid background for string theory.

The boundary CFT metric obtained as $\chi$ becomes positively infinite is
the $(n+1)$ dimensional flat spacetime metric written in unusual coordinates 
\begin{eqnarray}  \label{bou1}
ds^2=dr^2+r^2(-d\tau^2+\cosh^2 \tau d\Omega_{n-1}^2),
\end{eqnarray}
and is similar to the metric obtained as $\chi \to -\infty$ (the conformal
factor $1/r^4$ obtained in this case can be thrown away by a redefinition of
the "radial" coordinate). Calculation of the boundary stress tensor for the
spacetime (\ref{ads1}) is straightforward. If we assume that the AdS/CFT
correspondence can be extended to such asymptotically locally AdS spaces,
the dual description of this spacetime will be given by some sort of the SYM
theory on $R^{n+1}$.

Another interesting case is again the limit $Q^2=M^2$. For extremal RNdS
black holes in planar coordinates, a general analytic continuation yielding
Lorentzian AdS solutions with the right signature seems impossible. However $%
Q^2=M^2$ does not correspond to an extremal black hole when a cosmological
constant is present; hence, at least in four dimensions, we can generate new
solutions of the EM equations on the Euclidean section, with $\Lambda<0$. An
AdS counterpart for the KT solution (\ref{Nbh}) is obtained by using the
Wick rotations $t \to i \chi,~~H \to -i H$. Here we should consider
magnetically-charged black holes in order to find real solutions. This
corresponds to multi-center Einstein-Maxwell instantons in AdS$_{4}$.

The two-center solution takes a particularly simple form. We start by
writing the $N=2$ KT solution in a prolate spheroidal coordinate system \cite%
{c}. Without loss of generality, the masses are placed at $(0,0,1)$ and $%
(0,0,-1)$ respectively. Thus, by defining 
\begin{eqnarray}  \label{tds}
x^1=\sinh \psi \sin \theta \cos \varphi,~~ x^2=\sinh \psi \sin \theta \sin
\varphi,~~ x^1=\cosh \psi \cos \theta,
\end{eqnarray}
the two-center KT metric becomes 
\begin{eqnarray}  \label{2bh}
ds^2=-U^{-2}dt^2+e^{2Ht}U^2\big(P(d\psi^2+d\theta^2)+ \sinh^2 \psi \sin^2
\theta d\varphi^2\big ),
\end{eqnarray}
where 
\begin{eqnarray}
U=1+We^{-Ht}/P, ~~~P=(\sinh^2 \psi + \sin^2 \theta), ~~ W=(M_1+M_2)\cosh
\psi+(M_1-M_2) \cos \theta.  \nonumber
\end{eqnarray}
We consider here a purely magnetic potential $A_{\varphi}=4 \cos \theta
\sinh^2 \psi /(\cos 2\theta -\cosh 2 \psi)$. The corresponding two-center
Euclidean $AdS_4$ solution of EM equations is found by analytically
continuing $t \to i \chi,~~H \to -i H$ and reads 
\begin{eqnarray}  \label{2bh-2}
ds^2=(1+\frac{e^{-H\chi}W}{P})^{-2}d\chi^2+e^{2H\chi} (1+\frac{e^{-H\chi}W}{P%
})^2\Big(P(d\psi^2+d\theta^2)+ \sinh^2 \psi \sin^2 \theta d\varphi^2\Big ),
\end{eqnarray}
with $P,W,A_{\varphi}$ still given by the above relations. The boundary CFT
metric obtained in this case for large positive $\chi$ corresponds again to
an unusual parametrization of the flat space 
\begin{eqnarray}  \label{2bb-b1}
ds^2=P(d\psi^2+d\theta^2)+ \sinh^2 \psi \sin^2 \theta d\varphi^2 ,
\end{eqnarray}
while in the limit $\chi \to -\infty$ the above metric gains a conformal
factor $(W/P)^2$.

A more detailed discussion of these solutions will be presented elsewhere. 

\section{Discussion}

Unlike the AdS/CFT correspondence, the conjectured dS/CFT correspondence is
far from being understood. A first step toward this understanding is to
study various configurations in asymptotically dS space. To this end we have
studied some properties of the RNdS black hole and its generalization (KT
solution) and argued that these admit descriptions in terms of the dS/CFT
correspondence. Working on the gravity side we tried to get some insight on
the properties of the \textit{putative} dual CFT. The powerful tool that we
used to do this is the counterterm prescription \cite{Balasubramanian:2001nb}
for the renormalization of the field theory stress-tensor.

The AdS/CFT correspondence is a concrete realization of the holographic
principle. Such correspondence is referred to as duality in the sense that
the supergravity (closed string) description of D-branes and the field
theory (open string) description are different formulations of the same
physics. This way, the infrared (IR) divergences of quantum gravity in the
bulk are equivalent to ultraviolet (UV) divergences of dual field theory
living on the boundary. When we specify the CFT and say on which space it
lives we are implicitly providing a set of counterterms for the gravity
solution. These counterterms are local and depend only on the intrinsic
boundary geometry \cite{bala}.

In ref.~\cite{Strominger:2001pn}, Strominger proposed a dS/CFT duality: bulk
dS physics is dual to a boundary CFT (consequently, bulk quantum states are
dual to CFT states on the boundary) and bulk time translation is dual to the
boundary scale transformation. Since the bulk isometry group $SO(D,1)$
agrees with the boundary conformal group of a single Euclidean CFT in $D$
dimensions, it seems the dS/CFT correspondence involves a single dual field
theory\footnote{%
In ref.~\cite{vijay} it was suggested that the duality should involve two
CFTs and dS spacetime is defined as a correlated state in Hilbert space of
the two field theories.} (see, $e.g.$, ref.~\cite{rob} for a nice
discussion). In the counterterm prescription we should also specify the
surfaces on which the counterterms have to be integrated (equivalently, we
should choose the space over which the dual field theory is defined). Note
that these surfaces do not in general enclose anything --- instead the
conserved quantity is associated with the boundary surface itself. This
situation is analogous to that in asymptotically AdS or flat spacetimes;
although the boundary surfaces in these cases enclose a bulk spatial region,
the conserved quantities are completely insensitive to what transpires in
the bulk, provided no information travels from bulk to boundary (or
vice-versa) \cite{booth}.

We note that in general the counterterm method does not always work. For
example, in dS spacetime in global coordinates, the action is finite in even
spatial dimensions only up to a term that linearly diverges in the
cosmological time $\mathcal{T}$. Since all such boundary counterterm
invariants are independent of this quantity, they cannot render the action
finite \cite{Ghezelbash:2002ab}. This resembles the AdS/CFT case with two
disconnected boundaries. However, in dS case the boundaries $I^{+/-}$ are
causally connected and are two different slices of the same foliation.

Banks conjectured in ref.~\cite{banks} that a quantum theory of gravity wih $%
\Lambda>0$ has a finite number of degrees of freedom. The holographic screen
is at the cosmological horizon, that is the largest surface from which an
inertial observer can receive information. This proposal led to a conjecture
by Bousso \cite{Bousso:2000md}, referred to as the N-bound: ``any
asymptotically dS spacetime will have an entropy no greater than the entropy 
$\pi/ H ^{2}$ of pure dS with cosmological constant $\Lambda =3H^{2}$ in $%
(3+1)$ dimensions''. From this the authors of ref.~\cite{bala} were further
led to the conjecture that ``any asymptotically dS spacetime with mass
greater than dS has a cosmological singularity'' (see, $e.g.$, ref.~\cite%
{cai1}). Roughly speaking this latter conjecture implies that the conserved
mass of any physically reasonable asymptotically dS spacetime must be
negative ($i.e.$ less than the zero value of pure dS spacetime). It has been
shown that both of these conjectures can be violated for asymptotically dS
spacetimes with non-zero NUT charge \cite{Clarkson2003ab} (see, also, ref.~%
\cite{wolf}). However we have found that asymptotically dS spacetimes with a 
$U(1)$ charge respect both conjectures.

Many properties of RNdS black holes that are obtained from a static
coordinate system remain valid in a planar coordinate system that makes
evident the expansion of the universe. The counterterm prescription yields a
vanishing Casimir energy and the right expression for the gravitational
mass, while the entropy of the cosmological event horizon can be written in
a Cardy--Velinde form. The Cardy--Verlinde formula has recently been shown 
to be applicable for the entropy of topological black holes \cite{Setare}. Althoug 
this formula was verified in the AdS/CFT context also, its interpretation is 
unclear. The main reason is that modular invariance, which is at the base of 
proving the Cardy formula, is a characteristic feature of two-dimensional CFTs 
only.

If there is a dS/CFT correspondence one is tempted to interpret time
translation in the bulk as a boundary scale transformation \cite{strominger}%
. By generalizing this beyond pure dS, time can be holographically
reconstructed: time evolution in an asymptotically \textit{expanding} dS
space is equivalent to an $inverse$ RG flow. Using different foliations for
the RNdS spacetime, we obtain a time-dependent effective cosmological
constant associated with the slices of the foliation. In this way, we
provide nontrivial examples of the generalized dS $c$-theorem \cite{fred}:
``the effective cosmological constant is larger in the interior of the space
than at the (conformal) boundary''. The interpretation of this result is
less clear than in AdS/CFT case (see, $e.g$, ref.~\cite{fred} for a
discussion on ``accessible'' and ``available'' degrees of freedom on a given
time slice). In our case, it would be more natural to interpret the inverse
RG flows from a state with less entropy to a pure dS state. That is, the
black hole is evaporating and the area of the cosmological horizon is
increasing, approaching the area of a cosmological horizon of a pure dS
spacetime.

The planar coordinate system admits also multi-black hole solutions with
possible relevance for the cosmic censorship. An analysis of these
configurations similar to the single-black hole case seems difficult. 
The boundary stress tensor presents divergences, which can be interpreted as
reflecting the existence of the locations of the black hole sources in the
bulk theory.

The KT multi-black hole solution (\ref{Nbh}) has been the starting point in
a number of studies of cosmic censorship \cite{Brill:1993tm}. It would be
interesting to understand this better from the point of view of the
holographic theory. However, it seems to be rather difficult to propose a
consistent CFT description of this type of dS solutions. For example, let us
consider the case of two ``ingoing'' KT black holes with mass $M_{1}$ and $%
M_{2}$. At sufficiently early times, the holes are far apart (out of causal
contact) and near each one the metric approaches that of a single black
hole. Thus we have a state of approximate thermal equilibrium for every
black hole with its own dS horizon, at temperatures $T_{1}$ and $T_{2}$.
This is a consequence of the existence of approximate symmetries of the form
(\ref{gen}) in the region near each black hole. Later the black holes come
into causal contact and merge into a single black hole. The dS and black
hole horizons approach a state of thermal equilibrium at a common
temperature $T(M_{1}+M_{2},\Lambda )$. The topology of the boundary is
expected to change when the two black holes merge.

Here, the existence of an upper limit (Nariai mass) for the total mass
implies some strange results. As discussed in ref.~\cite{Brill:1993tm}, for $%
N=2$ colliding black holes that are each less than the extremal mass but
whose sum is greater, do produce naked singularities (we expect this result
to hold also for $N>2$). One can ask whether we can find a well-defined
Euclidean CFT dual to this situation. Late times in the bulk are supposed to
correspond to the UV of the boundary theory, while early times correspond to
the IR. Therefore we expect the dual theory to have highly non-equilibrated
IR states, whereas the naked singularity in the bulk will manifest itself in
UV pathologies of the CFT.

Let us take a closer look at this situation. There are some subtleties to
deal with before one can conclude that cosmic censorship is violated in the
KT solution. The naked singularity exists even before the black holes
collide, due to the singular initial conditions we choose. This problem can
be avoided by introducing a charged shell of dust that hides the naked
singularity. On the other hand there is always a generically singular Cauchy
horizon that has to be traversed by obsevers to see the naked singularity.
The black hole collision is only possible in the contracting patch ($H<0$),
since the KT solution describes masses at arbitrary positions but not
arbitrary velocities. If two undermassive black holes merge into an
overmassive one, an observer at rest in contracting patch has no way of
telling what is happening in the other ``half'' ($H>0$) of dS spacetime. She
only can guess the total mass will be $-(M_{1}+M_{2})$ to balance the mass
she sees. Crossing the cosmological horizon ($r=+\infty $), the masses will
appear to her very close together corresponding to very early time. However
it is reasonable to interpret the cosmological horizon discontinuities as
pulses of gravitational and electromagnetic radiation. It is clear now that
for generic boundaries conditions there will be gravitational waves
traveling along the Cauchy horizon\footnote{%
We thank Robert Myers for discussion on this point.}. When we give a
particular spacetime which is asymptotically dS, we are giving a dual CFT
living on the boundary and a particular state in the CFT. Matching smoothly
expanding and contracting patches across the cosmological horizon implies a
very precise selection for the CFT conditions ($i.e.$ very special geometric
data on the boundaries). It was shown in ref.~\cite{Brill:1993tm} (see,
also, ref.~\cite{brill}) that this happens in a very special situations:
several KT masses symmetrically distributed about a given one. If the
observer survives crossing the Cauchy horizon, then cosmic censorship is
violated in these examples\footnote{%
However, this is still not considered a serious violation of cosmic
censorship. It was argued in refs.~\cite{Brill:1993tm, brill} that if we
want to consider a more physical situation (to form black holes from
collapsing dust), the dust ball to form the black hole and the dust ball
that hides the ``overmassive'' singularity collide before any singularity
can form. Unlike the collision of two eternal black holes (for which the
initial data contains the infinite throats of the black holes), it is not
clear that starting with more generic initial data (compact data) the
singularity at the Cauchy horizon is not stronger than in the KT solution.}.

We ended with several remarks on a possible way to generate new solutions
with a negative cosmological constant starting with EM-dS configurations in
planar coordinates. We found that double analytic continuations of
coordinates of the class of metrics in planar coordinates yield a class of
bubble-like metrics given by eq.~(\ref{ads1}). These metrics are not bubbles
in the sense that there is no additional periodic coordinate outside of the
de Sitter subspace. However such metrics in principle furnish time-dependent
backgrounds for string theory --- their ultimate utility remains to be seen.

As a final comment we note that the counterterm method utilized here is a
perfectly valid tool for the regularization of infrared divergences and the
calculation of the gravitational mass of asymptotically dS spacetimes even
when removed from the context of the dS/CFT correspondence. Therefore, while
we are optimistic regarding this correspondence (however, see ref.~\cite%
{lisa}), the results described here hold independently of its validity. 

\subsection*{Acknowledgements}

DA would like to thank Robert Myers for bringing the refs.~\cite%
{Brill:1993tw,Brill:1993tm} to his attention, at the beginning of this work.
The authors would like to thank Robert Myers for valuable discussions and
David Winters for proof-reading an earlier draft of this paper and for help
with the figures. DA would like to thank the University of Waterloo's
Department of Physics, ICTP Trieste, UBC's Department of Physics and PIMS
Vancouver for their hospitality during various stages of this project.

The work of DA and RM was supported in part by the Natural Sciences and
Engineering Research Council of Canada. DA is further supported by a Dow
Hickson Fellowship. The work of ER was performed in the context of the
Graduiertenkolleg of the Deutsche Forschungsgemeinschaft (DFG): Nichtlineare
Differentialgleichungen: Modellierung, Theorie, Numerik, Visualisierung. 


\end{document}